\input phyzzx
\overfullrule=0pt
\def\sqr#1#2{{\vcenter{\hrule height.#2pt
      \hbox{\vrule width.#2pt height#1pt \kern#1pt
          \vrule width.#2pt}
      \hrule height.#2pt}}}

\def\square{{\mathchoice{\sqr84}{\sqr84}{\sqr{5.0}3}{\sqr{3.5}3}}}
\def\dA{\mathop\square\nolimits}
\def\Lambdato{{\buildrel{\Lambda\rightarrow\infty}\over=}}
\REF\HAY{%
T. Hayashi, Y. Ohshima, K. Okuyama and H. Suzuki, Prog.\ Theor.\
Phys.\ {\bf 100} (1998), 627.}
\REF\WES{%
J. Wess and J. Bagger, {\sl Supersymmetry and Supergravity}
(Princeton University Press, Princeton, 1992).}
\REF\GRI{%
M. T. Grisaru, W. Siegel and M. Ro\v cek, Nucl.\ Phys.\ {\bf B159}
(1979), 429.}
\REF\FUJ{%
K. Fujikawa, Phys.\ Rev.\ {\bf D29} (1984), 285.}
\REF\BAN{%
H. Banerjee, R. Banerjee and P. Mitra, Z. Phys.\ {\bf C32} (1986),
445.}
\REF\PIG{%
O. Piguet and K. Sibold, Nucl.\ Phys.\ {\bf B247} (1984), 484.}
\REF\CLA{%
T. E. Clark and S. Love, Phys.\ Lett.\ {\bf 138B} (1984), 289.}
\REF\NIE{%
N. K. Nielsen, Nucl.\ Phys.\ {\bf B244} (1984), 499.}
\REF\GUA{%
E. Guadagnini, K. Konishi and M. Mintchev, Phys.\ Lett.\ {\bf 157B}
(1985), 37.}
\REF\HAR{%
K. Harada and K. Shizuya, Phys.\ Lett.\ {\bf 162B} (1985), 322.}
\REF\NEM{%
D. Nemeschansky and R. Rohm, Nucl.\ Phys.\ {\bf B249} (1985), 157.}
\REF\GIR{%
G. Girardi, R. Grimm and R. Stora, Phys.\ Lett.\ {\bf 156B} (1985),
203.}
\REF\GAR{%
R. Garreis, M. Scholl and J. Wess, Z. Phys.\ {\bf C28} (1985), 628.}
\REF\KON{%
K. Konishi and K. Shizuya, Nuovo Cim.\ {\bf 90A} (1985), 111.}
\REF\FRO{%
S. A. Frolov and A. A. Slavnov, Phys.\ Lett.\ {\bf B309} (1993), 344.}
\REF\NAR{%
R. Narayanan and H. Neuberger, Phys.\ Lett.\ {\bf B302} (1993), 62.}
\REF\AOK{%
S. Aoki and Y. Kikukawa, Mod.\ Phys.\ Lett.\ {\bf A8} (1993), 3517.}
\REF\FUJI{%
K. Fujikawa, Nucl.\ Phys.\ {\bf B428} (1994), 169; Indian J.\ Phys.\
{\bf 70A} (1996), 275.}
\REF\OKU{%
K. Okuyama and H. Suzuki, Phys.\ Lett.\ {\bf B382} (1996), 117; Prog.\
Theor.\ Phys. {\bf 98} (1997), 463.}
\REF\CHA{%
L. N. Chang and C. Soo, Phys.\ Rev.\ {\bf D55} (1997), 2410.}
\REF\DEW{%
B. S. DeWitt, {\sl Dynamical Theory of Group and Fields} (Gordon and
Breach, New York, 1978).}
\REF\ABB{%
L. F. Abbott, M. T. Grisaru and R. K. Schaefer, Nucl.\ Phys.\
{\bf B229} (1983), 372.}
\REF\LUK{%
J. Lukierski, Phys.\ Lett.\ {\bf 70B} (1977), 183.}
\REF\CUR{%
T. Curtright, Phys.\ Lett.\ {\bf 71B} (1977), 185.}
\REF\INA{%
H. Inagaki, Phys.\ Lett.\ {\bf 77B} (1978), 56.}
\REF\LAN{%
W. Lang, Nucl.\ Phys.\ {\bf B150} (1979), 201.}
\REF\HAG{%
T. Hagiwara, S.-Y. Pi and H.-S. Tsao, Ann.\ of Phys.\ {\bf 130}
(1980), 282.}
\REF\NIC{%
H. Nicolai and P. K. Townsend, Phys.\ Lett.\ {\bf 93B} (1980), 111.}
\REF\PIGU{%
O. Piguet and K. Sibold, Nucl.\ Phys.\ {\bf B196} (1982), 447.}
\REF\MAR{%
S. Marculescu, Nucl.\ Phys.\ {\bf B213} (1983), 523.}
\REF\GRIS{%
M. T. Grisaru and P. T. West, Nucl.\ Phys.\ {\bf B254} (1985), 249.}
\REF\GRISA{%
M. T. Grisaru, B. Milewski and D. Zanon, Phys.\ Lett.\ {\bf 157B}
(1985), 174; Nucl.\ Phys.\ {\bf B266} (1986), 589.}
\REF\SHI{%
M. Shifman and A. Vainshtein, Nucl.\ Phys.\ {\bf B277} (1986), 456.}
\REF\MEH{%
M. R. Mehta, Phys.\ Rev.\ {\bf D44} (1991), 3303.}
\REF\FER{%
S. Ferrara and B. Zumino, Nucl.\ Phys.\ {\bf B87} (1975), 207.}
\REF\SHIZ{%
K. Shizuya, Phys.\ Rev.\ {\bf D35} (1987), 1848.}
\REF\ABBO{%
L. F. Abbott, Nucl.\ Phys.\ {\bf B185} (1981), 189.}
\REF\GAT{%
S. J. Gates, M. T. Grisaru, M. Ro\v cek and W. Siegel,
{\sl Superspace or One Thousand-and-One Lessons in Supersymmetry}
(Benjamin/Cummings, Reading, Mass., 1983).}
\REF\DVA{%
G. Dvali and M. Shifman, Phys.\ Lett.\ {\bf B396} (1997), 64.\nextline
A. Kovner, M. Shifman and A. Smilga, Phys.\ Rev.\ {\bf D56} (1997),
7978.\nextline
B. Chibisov and M. Shifman, Phys.\ Rev.\ {\bf D56} (1997), 7990.}
\REF\FUJIK{%
K. Fujikawa and K. Okuyama, Nucl.\ Phys.\ {\bf B521} (1998), 401.}
\pubnum={IU-MSTP/29; hep-th/9805068}
\date={May 1998}
\titlepage
\title{\hbox{Invariant Regularization of Supersymmetric Chiral Gauge
Theory. II}}
\author{%
Takuya Hayashi,\foot{e-mail: hayashi@physun1.sci.ibaraki.ac.jp}
Yoshihisa Ohshima\foot{e-mail: ohshima@mito.ipc.ibaraki.ac.jp}\break
Kiyoshi Okuyama\foot{e-mail: okuyama@mito.ipc.ibaraki.ac.jp}
and Hiroshi Suzuki\foot{e-mail: hsuzuki@mito.ipc.ibaraki.ac.jp}}
\address{%
Department of Physics, Ibaraki University, Mito 310, Japan}
\abstract{%
By undertaking additional analyses postponed in a previous paper,
we complete our construction of a manifestly supersymmetric
gauge-covariant regularization of supersymmetric chiral gauge
theories. We present the following: An evaluation of the covariant
gauge anomaly; a proof of the integrability of the covariant gauge
current in anomaly-free cases; a calculation of a one-loop
superconformal anomaly in the gauge supermultiplet sector. On the
last point, we find that the ghost-anti-ghost supermultiplet and the
Nakanishi-Lautrup supermultiplet give rise to BRST exact
contributions which, due to ``tree-level'' Slavnov-Taylor identities
in our regularization scheme, can safely be neglected, at least at
the one-loop level.}
\endpage
\chapter{Introduction}
In a recent paper~[\HAY], we proposed a manifestly supersymmetric
gauge-covariant regularization of supersymmetric {\it chiral\/} gauge
theories. In the sense of the background field method, it was shown
that our scheme provides a supersymmetric gauge invariant
regularization of the effective action {\it above\/} one-loop order.
On the other hand, our scheme gives a gauge covariant definition of
one-loop diagrams, and, when the representation of the chiral
supermultiplet is free from the gauge anomaly, the definition
restores the gauge invariance. Our scheme also defines composite
operators in a supersymmetric gauge-covariant manner. However, several
important issues concerning properties of our scheme were postponed
in Ref.~[\HAY]. In the present paper, to complete the construction, we
present detailed analyses on those issues.

We first recapitulate the essence of our regularization scheme.\foot{%
For conciseness, we do not repeat the explanation of our notation and
conventions given in Ref.~[\HAY]; these basically follow the
conventions of Ref.~[\WES].}
We consider a general renormalizable supersymmetric model:
$$
   S={1\over2T(R)}\int d^6z\,\tr W^\alpha W_\alpha
   +\int d^8z\,\Phi^\dagger e^V\Phi
   +\int d^6z\,\left({1\over2}\Phi^T m\Phi+{1\over3}g\Phi^3\right)
   +{\rm h.c.}
\eqn\onexone
$$
To apply the notion of the superfield background field method~[\GRI],
we split the gauge superfield and the chiral superfield as~[\HAY]
$$
   e^V=e^{V_B}e^{V_Q},\qquad\Phi=\Phi_B+\Phi_Q.
\eqn\onextwo
$$
Furthermore, to make the supersymmetry and the background gauge
invariance (in the unregularized level) manifest, we adapt the
following gauge fixing term and the ghost-anti-ghost term
$$
\eqalign{
   S'&=-{\xi\over8T(R)}\int d^8z\,\tr(\overline D^2V_Q)({\cal D}^2V_Q)
\cr
   &\qquad+{1\over T(R)}\int d^8z\,
   \tr(e^{-V_B}c^{\prime\dagger}e^{V_B}+c^\prime)
\cr
   &\qquad\qquad\qquad\times
   {\cal L}_{V_Q/2}\cdot
   \left[(c+e^{-V_B}c^\dagger e^{V_B})
    +\coth({\cal L}_{V_Q/2})\cdot
   (c-e^{-V_B}c^\dagger e^{V_B})\right]
\cr
   &\qquad-{2\xi\over T(R)}\int d^8z\,\tr e^{-V_B}b^\dagger e^{V_B}b,
\cr
}
\eqn\onexthree
$$
where the normalization of the Nielsen-Kallosh (NK) ghost~$b$ has
been changed from Ref.~[\HAY].

In calculating radiative corrections to the effective action in the
background field method, i.e., the generating functional of 1PI
Green's functions with all the external lines forming the background
field, $V_B$ or~$\Phi_B$, we expand the total action~$S_T\equiv S+S'$
in powers of the {\it quantum\/} field, $S_T=S_{T0}+S_{T1}+S_{T2}
+S_{T3}+\cdots$. (Hereafter, a number appearing in the subscript
indicates the power of the quantum fields.) The quadratic action is
further decomposed
as $S_{T2}=S_{T2}^{\rm gauge}+S_{T2}^{\rm ghost}+S_{T2}^{\rm chiral}
+S_{T2}^{\rm mix}$.

The first part, which is composed purely of the gauge superfields, is
given by
$$
\eqalign{
   &S_{T2}^{\rm gauge}
\cr
   &=\int d^8z\,V_Q^a
   \left[{1\over8}\widetilde\nabla^\alpha\overline D^2
         \widetilde\nabla_\alpha
         +{1\over2}{\cal W}_B^\alpha\widetilde\nabla_\alpha
         -{\xi\over16}(\widetilde\nabla^2\overline D^2
         +\overline D^2\widetilde\nabla^2)\right]^{ab}
   V_Q^b
\cr
   &=\int d^8z\,V_Q^a
   \left[-\widetilde\nabla^m\widetilde\nabla_m
   +{1\over2}{\cal W}_B^\alpha\widetilde\nabla_\alpha
   -{1\over2}\overline{\cal W}_{B\dot\alpha}^\prime
    \overline D^{\dot\alpha}
   +{1\over16}(1-\xi)(\widetilde\nabla^2\overline D^2
            +\overline D^2\widetilde\nabla^2)\right]^{ab}V_Q^b.
\cr
}
\eqn\onexfour
$$
The ghost-anti-ghost action, to second order in the quantum fields,
is given by
$$
   S_{T2}^{\rm ghost}
   =\int d^8z\,\left[c^{\prime\dagger a}(e^{{\cal V}_B})^{ab}c^b
                    +c^{\dagger a}(e^{{\cal V}_B})^{ab}c^{\prime b}
                 -2\xi b^{\dagger a}(e^{{\cal V}_B})^{ab}b^b\right].
\eqn\onexfive
$$
There are two kinds of action which contain the chiral multiplet. One
is the part that survives even for~$\Phi_B=0$,
$$
   S_{T2}^{\rm chiral}
   =\int d^8z\,\Phi_Q^\dagger e^{V_B}\Phi_Q
   +\int d^6z\,{1\over2}\Phi_Q^Tm\Phi_Q+{\rm h.c.},
\eqn\onexsix
$$
and the other is the part that disappears for~$\Phi_B=0$,
$$
\eqalign{
   S_{T2}^{\rm mix}
   &=\int d^8z\,\left(\Phi_B^\dagger e^{V_B}V_Q\Phi_Q
                     +\Phi_Q^\dagger e^{V_B}V_Q\Phi_B
            +{1\over2}\Phi_B^\dagger e^{V_B}V_Q^2\Phi_B\right)
\cr
   &\qquad+\int d^6z\,g\Phi_B\Phi_Q^2+{\rm h.c.}
\cr
}
\eqn\onexseven
$$

Our regularization is then implemented as follows: We take propagators
of the quantum fields that are given by formally diagonalizing
$S_{T2}^{\rm gauge}+S_{T2}^{\rm ghost}+S_{T2}^{\rm chiral}$.\foot{%
There is no deep reason for doing this. It merely simplifies the
expressions, because we do not include~$S_{T2}^{\rm mix}$ in
diagonalizing the quadratic action.}
Then, for a finite ultraviolet cutoff~$\Lambda$, we modify the
propagators so as to improve the ultraviolet behavior and
simultaneously preserve the background gauge covariance. For example,
for the quantum gauge superfield, we use\foot{%
Hereafter, the brackets~$\VEV{\cdots}$ represent an expectation value
in the unconventional perturbative picture in which
$S_{T2}^{\rm gauge}+S_{T2}^{\rm ghost}+S_{T2}^{\rm chiral}$ is
regarded as the ``un-perturbative part''.}
$$
\eqalign{
   &\VEV{T^*V_Q^a(z)V_Q^b(z')}
\cr
   &\equiv{i\over2}\Biggl[
   f\left([-\widetilde\nabla^m\widetilde\nabla_m
   +{\cal W}_B^\alpha\widetilde\nabla_\alpha/2
   -\overline{\cal W}_{B\dot\alpha}^\prime
    \overline D^{\dot\alpha}/2
   +(1-\xi)(\widetilde\nabla^2\overline D^2
            +\overline D^2\widetilde\nabla^2)/16]
   \bigm/(\xi\Lambda^2)\right)
\cr
   &\qquad\qquad
   \times{1\over-\widetilde\nabla^m\widetilde\nabla_m
   +{\cal W}_B^\alpha\widetilde\nabla_\alpha/2
   -\overline{\cal W}_{B\dot\alpha}^\prime
   \overline D^{\dot\alpha}/2
   +(1-\xi)(\widetilde\nabla^2\overline D^2
            +\overline D^2\widetilde\nabla^2)/16}
   \Biggr]^{ab}
\cr
   &\qquad\qquad\qquad\qquad\qquad\qquad\qquad\qquad\qquad
   \qquad\qquad\qquad\qquad\qquad\qquad
   \times\delta(z-z'),
\cr
}
\eqn\onexeight
$$
and, for the ghost superfields,
$$
\eqalign{
   &\VEV{T^*c^a(z)c^{\prime\dagger b}(z')}
   =\VEV{T^*c^{\prime a}(z)c^{\dagger b}(z')}
   =-2\xi\VEV{T^*b^a(z)b^{\dagger b}(z')}
\cr
   &\equiv
   i\left[
   f(-\overline D^2\widetilde\nabla^2/16\Lambda^2)
   \overline D^2{1\over\widetilde\nabla^2\overline D^2}
   \widetilde\nabla^2e^{-{\cal V}_B}\right]^{ab}
   \delta(z-z'),
\cr
}
\eqn\onexnine
$$
and, for the quantum chiral superfield,
$$
\eqalign{
   &\VEV{T^*\Phi_Q(z)\Phi_Q^\dagger(z')}
\cr
   &\equiv{i\over16}f(-\overline D^2\nabla^2/16\Lambda^2)
   \overline D^2
   {1\over\nabla^2\overline D^2/16-m^\dagger m}\nabla^2e^{-V_B}
   \delta(z-z'),
\cr
}
\eqn\onexten
$$
and
$$
   \VEV{T^*\Phi_Q(z)\Phi_Q^T(z')}
   \equiv{i\over4}f(-\overline D^2\nabla^2/16\Lambda^2)
   \overline D^2
   {1\over\nabla^2\overline D^2/16-m^\dagger m}m^\dagger
   \delta(z-z').
\eqn\onexeleven
$$
In these expressions, $f(t)$~is the regularization factor, which
decreases sufficiently rapidly, $f(\infty)=f'(\infty)=f''(\infty)
=\cdots=0$ in the ultraviolet, and $f(0)=1$, to reproduce the original
propagators in the infinite cutoff limit~$\Lambda\to\infty$. The
argument of the regularization factor has the same form as the
denominator of each propagator; this prescription is suggested by the
proper-time cutoff~[\HAY]. In this way, the propagators obey the same
transformation law as the original ones under the background gauge
transformation on the background gauge superfield~$V_B$. This
property is crucial for the gauge covariance of the scheme.

Using the above propagators of the quantum fields, 1PI Green's
functions are evaluated as follows. There are two kinds of
contributions, because we have diagonalized
$S_{T2}^{\rm gauge}+S_{T2}^{\rm ghost}+S_{T2}^{\rm chiral}$ in
constructing the propagators. (I)~Most of
radiative corrections are evaluated (as usual) by simply connecting
quantum fields in~$S_{T2}^{\rm mix}$, $S_{T3}$, $S_{T4}$, etc., by
the modified propagators. This defines the first part of the
effective action, ${\mit\Gamma}_{\rm I}[V_B,\Phi_B]$, which is given
by the 1PI part of
$$
   \VEV{\exp\left\{i\left[S_T
   -(S_{T2}^{\rm gauge}+S_{T2}^{\rm ghost}+S_{T2}^{\rm chiral})
   \right]\right\}}.
\eqn\onextwelve
$$
(II)~However, since the quadratic action~$S_{T2}^{\rm gauge}
+S_{T2}^{\rm ghost}+S_{T2}^{\rm chiral}$ depends on the background
gauge superfield~$V_B$ (but not on~$\Phi_B$) non-trivially, the
one-loop Gaussian determinant arising from this action has to be
taken into account. To {\it define\/} this part of the effective
action, ${\mit \Gamma}_{\rm II}[V_B]$, we adopt the following
prescription:
$$
   {\delta{\mit\Gamma}_{\rm II}[V_B]\over\delta V_B^a(z)}
   \equiv\VEV{J^a(z)},\qquad
   J^a(z)\equiv{\delta\over\delta V_B^a(z)}
   (S_{T2}^{\rm gauge}+S_{T2}^{\rm ghost}+S_{T2}^{\rm chiral}).
\eqn\onexthirteen
$$
The quantum fields in~$J^a(z)$ are connected by the modified
propagators. Note that since $J^a(z)$~is quadratic in the quantum
fields by definition, \onexthirteen~consists of {\it only} one-loop
diagrams. The total effective action is then given by the sum
${\mit \Gamma}[V_B,\Phi_B]={\mit \Gamma}_{\rm I}[V_B,\Phi_B]
+{\mit \Gamma}_{\rm II}[V_B]$. Although naively the
relation~\onexthirteen\ is just one of many equivalent ways to define
the one-loop effective action, it has a great advantage at the
regularized level; the prescription~\onexthirteen\ respects the gauge
covariance of the gauge current~[\HAY]. Our
prescription~\onexthirteen\ is a natural supersymmetric
generalization of the covariant regularization of Refs.~[\FUJ]
and~[\BAN]. (III)~When a certain composite operator~$O(z)$ is
inserted into a Green's function, it is computed as usual (by using
the modified propagators):
$$
   \VEV{O(z)\exp\left\{i\left[S_T
   -(S_{T2}^{\rm gauge}+S_{T2}^{\rm ghost}+S_{T2}^{\rm chiral})
   \right]\right\}}.
\eqn\onexfourteen
$$

Following the above prescription, it was shown~[\HAY] that the first
part of the effective action~${\mit\Gamma}_{\rm I}[V_B,\Phi_B]$
(which contains all the higher loop diagrams!) is always
supersymmetric and background gauge invariant. It was also shown that
a composite operator which behaves classically as a gauge covariant
superfield, such as the gauge current superfield~$\VEV{J^a(z)}$
in~\onexthirteen, is regularized as a (background) gauge covariant
superfield. This implies, in particular, that {\it if\/} there exists
a functional~${\mit\Gamma}_{\rm II}[V_B]$ in~\onexthirteen\ that
reproduces the gauge current superfield~$\VEV{J^a(z)}$ as the
variation, it is also supersymmetric and gauge invariant (because
$V_B^a$~is a superfield). One might expect that such a functional,
an ``effective action,'' always exists. However, this is not the case
in our scheme. In fact, if the whole effective action were always
gauge invariant, there would be no possibility of the gauge anomaly
that may arise from chiral multiplet's loop.

Actually, it is easy to see that the anomaly cancellation is the
{\it necessary\/} condition for the existence
of~${\mit\Gamma}_{\rm II}[V_B]$ in~\onexthirteen. As should be the
case, our scheme provides gauge invariant regularization {\it only\/}
for anomaly-free cases. In the analysis of Ref.~[\HAY], however, it
was not clear if the anomaly cancellation is also {\it sufficient\/}
for the existence of~${\mit\Gamma}_{\rm II}[V_B]$. In \S~2, we
establish the existence of~${\mit \Gamma}_{\rm II}[V_B]$ in
anomaly-free cases. Therefore, our scheme actually provides for
anomaly-free cases a gauge invariant regularization of the effective
action.

The prescription~\onexfourteen, on the other hand, is especially
useful when one wishes to evaluate the quantum anomaly while
preserving the supersymmetry and the gauge covariance (or
invariance). As was announced in Ref.~[\HAY], we apply it to the
superconformal anomaly in the gauge supermultiplet sector and present
the detailed one-loop calculation in \S~4. In this calculation,
the gauge-fixing term and the ghost-anti-ghost sector cause another
complication. To treat this complication systematically, in \S~3, we
introduce the notion of BRST symmetry and the corresponding
Slavnov-Taylor identity in our scheme.

\chapter{Integrability of the covariant gauge current}
As was noted in the Introduction, the most important issue that was
left uninvestigated in Ref.~[\HAY] is whether the second part of the
effective action~${\mit\Gamma}_{\rm II}[V_B]$ in~\onexthirteen\
exists; we call this a problem of the integrability of the gauge
current. The gauge anomaly cancellation~$\tr T^a\{T^b,T^c\}=0$ is the
necessary condition for the integrability~[\HAY] because the
covariantly regularized gauge current~$\VEV{J^a(z)}$ gives rise to
the covariant anomaly (we shall explicitly show this below), but, on
the other hand, $\VEV{J^a(z)}$ should produce a consistent
anomaly~[\PIG--\GAR] if such a
functional~${\mit\Gamma}_{\rm II}[V_B]$ exists. These two
requirements are consistent only if the gauge anomaly vanishes. In
this section, we shall establish the converse: When
$\tr T^a\{T^b,T^c\}=0$, ${\mit\Gamma}_{\rm II}[V_B]$ which
satisfies~\onexthirteen\ {\it always\/} exists
in~$\Lambda\to\infty$. Namely, the anomaly cancellation is also a
sufficient condition for the integrability. Therefore, when the gauge
representation is anomaly-free, our prescription~\onexthirteen\
provides a gauge invariant regularization of the effective action. We
shall give two proofs from quite different viewpoints. However,
before going into the proof, let us show how the gauge anomaly is
evaluated in our scheme.

\section{Covariant gauge anomaly}
In this subsection, we present a calculation of the covariant gauge
anomaly which reads\foot{%
This result itself is not new~[\KON]; we present the calculation in
our scheme for later use.}
$$
   -{1\over4}\overline D^2C^{ab}\VEV{J^b(z)}
   \Lambdato-{1\over64\pi^2}\tr T^aW_B^\alpha W_{B\alpha},
\eqn\twoxone
$$
where the combination~$C^{ab}$ has been defined by
$$
   C^{ab}
   \equiv\left[{{\cal V}_B\over2}
         \left(\coth{{\cal V}_B\over2}-1\right)\right]^{ab},
   \qquad{\cal V}_B\equiv{\cal T}^aV_B^a.
\eqn\twoxtwo
$$
The gauge covariance of the right-hand side of~\twoxone\ is in accord
with our gauge covariant definition of the gauge current~[\HAY]. To
see how the left-hand side of~\twoxone\ is related to the ``gauge
anomaly,'' i.e., the non-invariance of the effective
action~${\mit\Gamma}_{\rm II}[V_B]$ under the background gauge
transformation,\foot{%
Recall that the first part of the effective
action~${\mit\Gamma}_{\rm I}[V_B,\Phi_B]$ is always gauge invariant
in our scheme.}
we introduce the generator of the {\it background\/} transformation:
$$
   G^a(z)
   \equiv-{1\over4}\overline D^2C^{ab}{\delta\over\delta V_B^b(z)}.
\eqn\twoxthree
$$
In fact, it is easy to see that a variation of an arbitrary
functional~$F[V_B]$ of~$V_B$ under the background transformation (see
(2.6) of Ref.~[\HAY])
$$
   \delta V_B
   =i{\cal L}_{V_B/2}\cdot
   \left[(\Lambda+\Lambda^\dagger)+\coth({\cal L}_{V_B/2})\cdot
   (\Lambda-\Lambda^\dagger)\right],
\eqn\twoxfour
$$
can be written as
$$
   \delta F[V_B]=\int d^6z\,i\Lambda^a(z)G^a(z)F[V_B]+{\rm h.c.}
\eqn\twoxfive
$$
Therefore, assuming the identification in~\onexthirteen, the gauge
variation may be written as
$$
\eqalign{
   \delta{\mit\Gamma}_{\rm II}[V_B]
   &=\int d^6z\,i\Lambda^a(z)G^a(z){\mit\Gamma}_{\rm II}[V_B]
   +{\rm h.c.}
\cr
   &=\int d^6z\,i\Lambda^a(z)\left(-{1\over4}\right)
   \overline D^2C^{ab}\VEV{J^b(z)}+{\rm h.c.}
\cr
}
\eqn\twoxsix
$$
This is the reason we regard~\twoxone\ as the gauge anomaly.

Now, from the definition of the gauge current in~\onexthirteen, we
have
$$
\eqalign{
   &\int d^6z\,i\Lambda^a(z)\left(-{1\over4}\right)
   \overline D^2C^{ab}\VEV{J^b(z)}
\cr
   &=\int d^6z\,\VEV{i\Lambda^a(z)G^a(z)
   (S_{T2}^{\rm gauge}+S_{T2}^{\rm ghost}+S_{T2}^{\rm chiral})}.
\cr
}
\eqn\twoxseven
$$
However, $S_{T2}^{\rm gauge}$, $S_{T2}^{\rm ghost}$
and~$S_{T2}^{\rm chiral}$ are unchanged if we make the background
gauge transformation~\twoxfour\ and {\it simultaneously}
$$
   \delta V_Q^a=-i\Lambda^c({\cal T}^c)^{ab}V_Q^b,\qquad
   \delta\Phi_Q=-i\Lambda^aT^a\Phi_Q,
\eqn\twoxeight
$$
because $S_{T2}^{\rm gauge}$, $S_{T2}^{\rm ghost}$
and~$S_{T2}^{\rm chiral}$ are background gauge invariant. Therefore,
the right-hand side of~\twoxseven\ is equal to the {\it opposite\/}
of the variation of $S_{T2}^{\rm gauge}$, $S_{T2}^{\rm ghost}$
and~$S_{T2}^{\rm chiral}$ under~\twoxeight. This yields (in the
super-Fermi-Feynman gauge~$\xi=1$)
$$
\eqalign{
   &\int d^6z\,\VEV{i\Lambda^a(z)G^a(z)
   (S_{T2}^{\rm gauge}+S_{T2}^{\rm ghost}+S_{T2}^{\rm chiral})}
\cr
   &=-2i\int d^8z\,\Lambda^a(z)({\cal T}^a)^{bc}
   \VEV{V_Q^b\left(-\widetilde\nabla^m\widetilde\nabla_m
   +{1\over2}{\cal W}_B^\alpha\widetilde\nabla_\alpha
   -{1\over2}\overline{\cal W}_{B\dot\alpha}^\prime
    \overline D^{\dot\alpha}\right)^{cd}
   V_Q^d(z)}
\cr
   &\qquad+i\int d^8z\,\Lambda^a(z)({\cal T}^a)^{bc}
   \biggl[\VEV{c^{\prime\dagger d}(e^{{\cal V}_B})^{db}c^c(z)}
   +\VEV{c^{\dagger d}(e^{{\cal V}_B})^{db}c^{\prime c}(z)}
\cr
   &\qquad\qquad\qquad\qquad\qquad\qquad\qquad\qquad\qquad
   \qquad\qquad\quad
   -2\VEV{b^{\dagger d}(e^{{\cal V}_B})^{db}b^c(z)}\biggr]
\cr
   &\qquad+i\int d^8z\,
   \Lambda^a(z)\VEV{\Phi_Q^\dagger e^{V_B}T^a\Phi_Q(z)}
   +i\int d^6z\,
   \Lambda^a(z)\VEV{\Phi_Q^TmT^a\Phi_Q(z)},
\cr
}
\eqn\twoxnine
$$
where we have used the explicit form of the action given by
\onexfour--\onexsix. In the above expression, use of the modified
propagators \onexeight--\onexeleven\ is
assumed according to our prescription~\onexthirteen. Then the kinetic
operators in $S_{T2}^{\rm gauge}$, $S_{T2}^{\rm ghost}$
and~$S_{T2}^{\rm chiral}$ cancel the denominators of the propagators.
As a result, from~\twoxseven, we have
$$
\eqalign{
   &-{1\over4}\overline D^2C^{ab}\VEV{J^b(z)}
\cr
   &={i\over4}\overline D^2\lim_{z'\rightarrow z}
   \left\{{\cal T}^a
   f\left((-\widetilde\nabla^m\widetilde\nabla_m
   +{\cal W}_B^\alpha\widetilde\nabla_\alpha/2
   -\overline{\cal W}_{B\dot\alpha}^\prime
    \overline D^{\dot\alpha}/2)
   /\Lambda^2\right)\right\}^{bb}\delta(z-z')
\cr
   &\qquad+3\cdot{i\over4}\lim_{z'\rightarrow z}
   \left[{\cal T}^a
   f(-\overline D^2\widetilde\nabla^2/16\Lambda^2)\right]^{bb}
   \overline D^2\delta(z-z')
\cr
   &\qquad-{i\over4}\lim_{z'\rightarrow z}
   \tr T^af(-\overline D^2\nabla^2/16\Lambda^2)\overline D^2
   \delta(z-z').
\cr
}
\eqn\twoxten
$$
The evaluation of this expression is straightforward. First of all,
the first term on the right-hand side vanishes as~$\Lambda\to\infty$
because the number of spinor derivatives is not sufficient; the gauge
multiplet does not contribute. Next, in the second (the
ghost-anti-ghost multiplet) and the third (the chiral multiplet)
terms, we can repeat the calculation in the Appendix of Ref.~[\HAY].
In this way, we finally obtain~\twoxone. Note that the
ghost-anti-ghost multiplet does not contribute, because
$\tr{\cal T}^a\{{\cal T}^b,{\cal T}^c\}=0$.

\section{Integrability of the gauge current: the first proof}
We prove that the second part of the effective
action~${\mit\Gamma}_{\rm II}[V_B]$ in~\onexthirteen\ always exists
when the representation of the chiral multiplet is free of the gauge
anomaly. Our first proof is a natural supersymmetric generalization
of the procedure in Ref.~[\BAN], which starts with the {\it answer}
$$
   {\mit\Gamma}_{\rm II}[V_B]=\int_0^1dg\int d^8z\,
   V_B^a(z)\VEV{J^a(z)}_g.
\eqn\twoxeleven
$$
In this expression, the subscript~$g$ implies the expectation value
is evaluated with the ``coupling constant'' introduced
by taking~$V_B\to gV_B$ in our definition of the covariant gauge
current~\onexthirteen. We show below that, when
$\tr T^a\{T^b,T^c\}=0$, this functional reproduces the gauge current
in the limit~$\Lambda\to\infty$,
$$
   {\delta{\mit\Gamma}_{\rm II}[V_B]\over\delta V_B^a(z)}
   =\VEV{J^a(z)}.
\eqn\twoxtwelve
$$
Note that, since the composite operator~$\VEV{J^a(z)}_g$
in~\twoxeleven\ does depend on the background gauge field~$V_B$, the
relation~\twoxtwelve\ is by no means trivial.

We now proceed with the proof. We first directly take the functional
derivative of~${\mit\Gamma}_{\rm II}[V_B]$~\twoxeleven:
$$
\eqalign{
   {\delta{\mit\Gamma}_{\rm II}[V_B]\over\delta V_B^a(z)}
   &=\int_0^1dg\,\VEV{J^a(z)}_g
   +\int_0^1dg\int d^8z'\,
   V_B^b(z'){\delta\VEV{J^b(z')}_g\over\delta V_B^a(z)}
\cr
   &=\VEV{J^a(z)}
   -\int_0^1dg\,g{d\over dg}\VEV{J^a(z)}_g
   +\int_0^1dg\int d^8z'\,
   V_B^b(z'){\delta\VEV{J^b(z')}_g\over\delta V_B^a(z)}.
\cr
}
\eqn\twoxthirteen
$$
In going from the first line to the second line, we performed an
integration by parts. Then we note $\VEV{J^a(z)}_g$~depends on~$g$
only through the combination~$gV_B$. Therefore
$$
   {d\over dg}\VEV{J^a(z)}_g
   =\int d^8z'\,
   V_B^b(z'){\delta\VEV{J^a(z)}_g\over g\delta V_B^b(z')}.
\eqn\twoxfourteen
$$
Substituting~\twoxfourteen\ into~\twoxthirteen, we have
$$
   {\delta{\mit\Gamma}_{\rm II}[V_B]\over\delta V_B^a(z)}
   =\VEV{J^a(z)}
   +\int_0^1dg\int d^8z'\,V_B^b(z')
   \left[{\delta\VEV{J^b(z')}_g\over\delta V_B^a(z)}
        -{\delta\VEV{J^a(z)}_g\over\delta V_B^b(z')}\right].
\eqn\twoxfifteen
$$
The relation~\twoxtwelve\ follows if the quantity in the square
brackets (the functional rotation) vanishes. To consider this
quantity, we note the identity
$$
   -{1\over4}\overline D^2C^{ac}
   {\delta\VEV{J^c(z)}\over\delta V_B^b(z')}
   ={\delta\over\delta V_B^b(z')}
   \left(-{1\over4}\right)\overline D^2C^{ac}\VEV{J^c(z)}
   +{1\over4}\overline D^2{\delta C^{ac}\over\delta V_B^b(z')}
   \VEV{J^c(z)}.
\eqn\twoxsixteen
$$
Note that the first term on the right-hand side is nothing but the
gauge anomaly~\twoxone. On the other hand, it is not difficult to see
that the gauge covariance of our gauge current~[\HAY],
$$
   \VEV{J^a(z)}'
   ={\partial V_B^b(z)\over\partial V_B'^a(z)}\VEV{J^b(z)},
\eqn\twoxseventeen
$$
where $V_B'=V_B+\delta V_B$ and $\delta V_B$ is given by~\twoxfour,
implies
$$
   -{1\over4}\overline D^2C^{ac}
   {\delta\VEV{J^b(z')}\over\delta V_B^c(z)}
   ={1\over4}\overline D^2{\delta C^{ac}\over\delta V_B^b(z')}
   \VEV{J^c(z)}
\eqn\twoxeighteen
$$
as a coefficient of the chiral gauge parameter~$\Lambda$. Therefore,
{}from \twoxeighteen, \twoxsixteen\ and~\twoxone, we find
$$
\eqalign{
   -{1\over4}\overline D^2C^{ac}
   \left[{\delta\VEV{J^b(z')}\over\delta V_B^c(z)}
        -{\delta\VEV{J^c(z)}\over\delta V_B^b(z')}\right]
   &\Lambdato{\delta\over\delta V_B^b(z')}
   \left({1\over64\pi^2}\tr T^aW_B^\alpha W_{B\alpha}\right)
\cr
   &=0,\qquad\hbox{if $\tr T^a\{T^b,T^c\}=0$}.
\cr
}
\eqn\twoxnineteen
$$
Similarly, by repeating the above procedure for the anti-chiral
part~$\Lambda^\dagger$, we have
$$
   -{1\over4}D^2C^{ac}(V_B\to-V_B)
   \left[{\delta\VEV{J^b(z')}\over\delta V_B^c(z)}
        -{\delta\VEV{J^c(z)}\over\delta V_B^b(z')}\right]
   \Lambdato0,
\eqn\twoxtwenty
$$
if~$\tr T^a\{T^b,T^c\}=0$. To see the functional rotation in the
square brackets itself vanishes in \twoxnineteen\ and~\twoxtwenty, we
expand the functional rotation in powers of~$V_B$, as
$c_0+c_1V_B+c_2V_B^2+\cdots$. Then \twoxnineteen\ and~\twoxtwenty\
require $\overline D^2c_0=D^2c_0=0$, as the $O(V_B^0)$~term
(note that $C^{ab}=\delta^{ab}+O(V_B)$) and thus
$(\dA+iD\sigma^m\overline D\partial_m/2)c_0=0$. Assuming the boundary
condition that the functional rotation vanishes as
$x-x'\to\infty$,\foot{%
Otherwise, the integral in~\twoxfifteen, and thus \twoxeleven, would
be ill-defined.} we have~$c_0=0$. This procedure can clearly be
repeated for $c_1$, $c_2$, etc., and we finally conclude that $c_0
=c_1=\cdots=0$; that is, the functional rotation itself vanishes.
Therefore, from~\twoxfifteen, we have~\twoxtwelve.

Let us summarize what we have shown. The variation of the
functional~\twoxeleven\ reproduces the covariant gauge current
as~\twoxtwelve\ if the ``gauge anomaly,'' the {\it left-hand\/} side
of~\twoxone, vanishes. In the $\Lambda\to\infty$ limit, the ``gauge
anomaly'' is given by the {\it right-hand\/} side of~\twoxone\ which
vanishes when the anomaly cancellation condition,
$\tr T^a\{T^b,T^c\}=0$, holds. Therefore, if~$\tr T^a\{T^b,T^c\}=0$,
the effective action~\twoxeleven\ satisfies \twoxtwelve\ in the
$\Lambda\to\infty$~limit. Put differently, even
if~$\tr T^a\{T^b,T^c\}=0$, when the cutoff~$\Lambda$ is finite, there
may exist pieces in the covariantly regularized gauge
current~$\VEV{J^a(z)}$ which cannot be expressed as a variation of
some functional. However, those non-integrable pieces disappear in
the limit $\Lambda\to\infty$ in the same sense that the gauge anomaly
is given by the right-hand side of~\twoxone\ in this same limit.

\section{Integrability of the gauge current: the second proof}
In this subsection, we give another proof of the integrability. This
proof, although being somewhat less rigorous and not applicable to
the general form of the regularization factor~$f(t)$, demonstrates an
interesting relation between our prescription~\onexthirteen\ and the
generalized Pauli-Villars regularization~[\FRO--\CHA]. We show below
that, when the gauge anomaly vanishes, \onexthirteen\ can basically
be realized by the generalized Pauli-Villars regularization. Since
the Pauli-Villars regularization is implemented at the level of
Feynman diagrams for which the Bose symmetry among gauge vertices is
manifest, the corresponding effective action always exists; namely,
the integrability is obvious. Since the gauge multiplet and the
ghost-anti-ghost multiplet belong to the adjoint (i.e., real)
representation, one may always apply the conventional Pauli-Villars
prescription to those sectors; there is no subtlety associated with
the gauge anomaly. Therefore, we shall present only analysis on the
chiral multiplet.

To relate our prescription with the generalized Pauli-Villars
regularization for the complex gauge representation, we introduce the
``doubled'' representation following Ref.~[\CHA] (see also
Ref.~[\OKU]):
$$
   R^a\equiv\pmatrix{T^a&0\cr0&-T^{aT}\cr},\qquad
   U_B\equiv R^aV_B^a,
\eqn\twoxtwentyone
$$
where $T^a$~is the original (complex) gauge representation. Then,
with this notation, our regularized gauge current~\onexthirteen\
for the chiral multiplet is expressed as
$$
\eqalign{
   &\VEV{J_{\rm chiral}^a(z)}
   \equiv\VEV{{\delta S_{T2}^{\rm chiral}\over\delta V_B^a(z)}}
\cr
   &={i\over16}\lim_{z\rightarrow z'}
   \tr{1+\sigma^3\over2}
   e^{-{U_B}}{\partial e^{U_B}\over\partial V_B^a(z)}
   f(-\overline D^2\nabla^2/16\Lambda^2)
   \overline D^2
   {1\over\nabla^2\overline D^2/16-m^2}\nabla^2
   \delta(z-z').
\cr
}
\eqn\twoxtwentytwo
$$
Throughout this subsection, all the covariant derivatives are
understood as the ``doubled'', $\nabla_\alpha
\equiv e^{-{U_B}}D_\alpha e^{U_B}$. In~\twoxtwentytwo, $\sigma^3$~is
the Pauli matrix in the doubled space, and we have inserted the
projection operator~$(1+\sigma^3)/2$ to extract the original
representation~$T^a$ from the doubled representation. Since
$\sigma^3R^a=R^a\sigma^3$, the position of the projection operator
does not matter.

We shall show that if $\tr T^a=0$ and $\tr T^a\{T^b,T^c\}=0$, there
exists a Lagrangian of Pauli-Villars regulators which basically
reproduces our prescription~\twoxtwentytwo. To write down this
Lagrangian, we introduce a set of infinite regulators~[\FRO]
$\phi_j$ ($j=0$, $1$, $2$, $\cdots$). We assume $\phi_j$ with $j$
even is a Grassman-even chiral superfield and that with $j$ odd is a
Grassman-odd chiral superfield. Then the action is given by
$$
\eqalign{
   S_{\rm PV}
   &=\sum_{j=0,2,4,\cdots}
   \left(
   \int d^8z\,\phi_j^\dagger e^{U_B}\phi_j
   +\int d^6z\,{1\over2}M_j\phi_j^T\sigma^1\phi_j+{\rm h.c.}
   \right)
\cr
   &\qquad+\sum_{j=1,3,5,\cdots}
   \left(
   \int d^8z\,\phi_j^\dagger\sigma^3e^{U_B}\phi_j
   +\int d^6z\,{1\over2}M_j\phi_j^Ti\sigma^2\phi_j+{\rm h.c.}
   \right),
\cr
}
\eqn\twoxtwentythree
$$
where $M_0$ is the original mass, $M_0=m$, and the mass of regulators
is assumed to be of the order of the cutoff, $M_j=n_j\Lambda$
for~$j\neq0$. Note that the use of the Pauli matrices $\sigma^1$
and~$i\sigma^2$ makes possible the existence of mass that are
consistent with the statistics of the fields. Noting the relations
$$
   \sigma^1R^a=-R^{aT}\sigma^1,\qquad
   i\sigma^2R^a=-R^{aT}i\sigma^2,
\eqn\twoxtwentyfour
$$
it is easy to verify that \twoxtwentythree~is {\it invariant\/} under
the background gauge transformation. Therefore, the generalized
Pauli-Villars regularization cannot apply to the evaluation of the
gauge anomaly, and it is workable only for anomaly-free cases. The
regularization proceeds as follows~[\FRO]: The zeroth field~$\phi_0$
is introduced so as to simulate the one-loop contribution of the
original field~$\Phi_Q$. Therefore, the gauge representation~$R^a$
must be projected on the original representation~$T^a$ by
inserting~$(1+\sigma^3)/2$ into a certain point of the loop diagram.
The position and the number of the insertion do not matter, because
the free propagator~$\VEV{T^*\phi_0(z)\phi_0^\dagger(z')}_0$ and the
gauge generator~$R^a$ commute with~$\sigma^3$. On the other hand,
contributions of regulators~$\phi_j$ are simply summed over without
the projection.\foot{%
Of course, for the gauge invariance, the momentum assignment for all
the fields must be taken the same.}

It is quite convenient to re-express the above prescription in terms
of the gauge current operator~[\FUJI--\CHA]. According to the
above prescription, the gauge current of the zeroth field~$\phi_0$ is
defined by
$$
   \VEV{J_0^a(z)}
   \equiv{i\over16}\lim_{z\rightarrow z'}
   \tr{1+\sigma^3\over2}
   e^{-{U_B}}{\partial e^{U_B}\over\partial V_B^a(z)}
   \overline D^2
   {1\over\nabla^2\overline D^2/16-m^2}\nabla^2
   \delta(z-z'),
\eqn\twoxtwentyfive
$$
where we have used the formal form of the propagator of~$\phi_0$ in
the background gauge field. The gauge current of the regulators is
similarly given by
$$
   \VEV{J_{j\neq0}^a(z)}
   ={i\over16}\lim_{z\rightarrow z'}
   e^{-{U_B}}{\partial e^{U_B}\over\partial V_B^a(z)}
   \overline D^2
   {(-1)^j\over\nabla^2\overline D^2/16-n_j^2\Lambda^2}\nabla^2
   \delta(z-z'),
\eqn\twoxtwentysix
$$
where the statistics of each field have been taken into account. The
sum of \twoxtwentyfive\ and~\twoxtwentysix\ gives the gauge current
operator in the generalized Pauli-Villars regularization:
$$
\eqalign{
   &\sum_{j=0,1,2,\cdots}\VEV{J_j^a(z)}
\cr
   &={i\over16}\lim_{z\rightarrow z'}
   \tr{1\over2}
   e^{-{U_B}}{\partial e^{U_B}\over\partial V_B^a(z)}
   f(-\overline D^2\nabla^2/16\Lambda^2)
   \overline D^2
   {1\over\nabla^2\overline D^2/16-m^2}\nabla^2
   \delta(z-z')
\cr
   &\qquad+{i\over16}\lim_{z\rightarrow z'}
   \tr{\sigma^3\over2}
   e^{-{U_B}}{\partial e^{U_B}\over\partial V_B^a(z)}
   \overline D^2
   {1\over\nabla^2\overline D^2/16-m^2}\nabla^2
   \delta(z-z').
\cr
}
\eqn\twoxtwentyseven
$$
In the above expression, we have introduced the function
$$
   f(t)=1+2\sum_{j=1}^\infty{(-1)^j\,t\over t+n_j^2}
\eqn\twoxtwentyeight
$$
by omitting an irrelevant term for~$m\ll\Lambda$. The function~$f(t)$
can be regarded as a regularization factor; in fact, one can
make~$f(t)$ sufficiently rapidly decreasing by choosing a suitable
sequence of regulator masses. For example, when~$n_j=j$~[\FRO], we
have
$$
   f(t)={\pi\sqrt{t}\over\sinh(\pi\sqrt{t})}.
\eqn\twoxtwentynine
$$
By comparing \twoxtwentyseven\ with~\twoxtwentytwo, we realize the
following correspondence: The first part of~\twoxtwentyseven\ is
identical to the~$1/2$ projected part of~\twoxtwentytwo, although the
regularization factor~$f(t)$ is limited to the form
of~\twoxtwentyeight. On the other hand, the second part
of~\twoxtwentyseven, the so-called ``parity-odd'' term, has no
regularization factor. Therefore the second part of~\twoxtwentyseven\
is UV divergent and ill-defined in general. However, it is known in
non-supersymmetric cases~[\FRO--\CHA] that when $\tr T^a
=\tr T^a\{T^b,T^c\}=0$, coefficients of these divergent pieces in the
``parity-odd'' term vanish. To see this is also the case
in~\twoxtwentyseven, we first note that
$e^{-{U_B}}\partial e^{U_B}/\partial V_B^a$ is Lie algebra-valued
(i.e., it is proportional to~$R^b$). Then it is easy to see that, by
examining the expansion in powers of~$V_B$, the quadratic, linear and
logarithmic divergences are proportional to $\tr\sigma^3R^a
=2\tr T^a$, $\tr\sigma^3R^aR^b=0$, $\tr\sigma^3R^aR^bR^c
=\tr T^a\{T^b,T^c\}$, respectively. Therefore, if~$\tr T^a
=\tr T^a\{T^b,T^c\}=0$, the second part of~\twoxtwentyseven\ is also
finite under the prescription~[\FRO--\CHA] that the trace over the
gauge indices is taken prior to the momentum integration.

Allow us to summarize the above results: The $1/2$~part of our
prescription~\twoxtwentytwo\ is basically equivalent to the
$1/2$~part of the generalized Pauli-Villars regularization,
\twoxtwentyseven. The $\sigma^3/2$~part of~\twoxtwentytwo\ in general
does not correspond to the $\sigma^3/2$~part of~\twoxtwentyseven.
However, if $\tr T^a=\tr T^a\{T^b,T^c\}=0$, the $\sigma^3/2$~part
of~\twoxtwentyseven\ (and thus also the $\sigma^3/2$~part
of~\twoxtwentytwo) is finite. Therefore, we may take the
$\Lambda\to\infty$~limit in the $\sigma^3/2$~part of~\twoxtwentytwo\
which, since $f(0)=1$, reduces to~\twoxtwentyseven. This shows that,
if the gauge anomaly vanishes (and if~$\tr T^a=0$),\foot{%
One might wonder why the condition $\tr T^a=0$, which did not appear
in the previous subsection, is required here. The reason is that the
{\it finiteness\/} of the ``parity-odd'' part is a stronger condition
than the {\it integrability}. In fact, the $V_B$-independent piece
of~$\VEV{J^a(z)}$ in our scheme acquires a quadratic divergence,
$\int_0^\infty dt\,f(t)\tr T^a\Lambda^2/(16\pi^2)$~[\HAY]. This does
not spoil the integrability as this can be expressed as a variation
of the Fayet-Iliopoulos $D$-term. However, this tadpole diagram is not
regularized by the Pauli-Villars prescription in this subsection.
This is the reason that $\tr T^a=0$ is required in the generalized
Pauli-Villars regularization.}
our prescription~\onexthirteen\ in the $\Lambda\to\infty$~limit is
basically equivalent to the generalized Pauli-Villars regularization.
This proves the integrability. This conclusion is consistent with the
result of the previous subsection that the integrability is ensured
only for the~$\Lambda\to\infty$~limit, even
with~$\tr T^a\{T^b,T^c\}=0$.

\chapter{BRST symmetry and the Slavnov-Taylor identity}
It is possible to introduce the notion of BRST symmetry in the
superfield background field method. Since the $S$-matrix in the
background field method is given by tree diagrams made from the
effective action~[\DEW,\ABB], which has $V_B$ and~$\Phi_B$ as the
argument, the relevance of the BRST symmetry for the unitarity of
physical $S$-matrix is not necessarily clear in the present
context.\foot{%
On the other hand, the renormalizability of the effective action in
the background field method is automatic in our scheme, because it
is restricted by the background gauge invariance, instead of the BRST
symmetry.}
However, we see in the next section that the notion is in fact quite
useful in systematically treating the gauge fixing part and the
ghost-anti-ghost sector. In this section, with this application in
mind, we present the method for (at least partially) incorporating
the BRST symmetry in our scheme.

Unfortunately, the BRST symmetry is not manifest in our scheme, and it
appears that the BRST symmetry or the associated Slavnov-Taylor (ST)
identity has to be verified order by order. In the second half of
this section, we show that there exists a natural extension of our
prescription of \S~1 which ensures ST identities at least at the
tree level, under the classical equation of motion of the background
field. This is sufficient for the application in the next section,
where the superconformal anomaly in the gauge sector is evaluated to
one-loop accuracy.

For later use, we summarize here the explicit form of the first
three terms in the expansion of the classical action of the gauge
sector,
$$
   S^{\rm gauge}={1\over2T(R)}\int d^6z\,\tr W^\alpha W_\alpha,
\eqn\threexone
$$
in powers of the quantum field~$V_Q$:
$$
   S_0^{\rm gauge}
   ={1\over2T(R)}\int d^6z\,\tr W_B^\alpha W_{B\alpha},
\eqn\threextwo
$$
$$
   S_1^{\rm gauge}
   =-{1\over T(R)}\int d^8z\,\tr V_Q{\cal D}^\alpha W_{B\alpha}
   =-{1\over C_2(G)}\int d^8z\,
   ({\cal V}_Q{\cal D}^\alpha{\cal W}_{B\alpha})^{aa},
\eqn\threexthree
$$
and
$$
   S_2^{\rm gauge}=\int d^8z\,V_Q^a
   \left({1\over8}\widetilde\nabla^\alpha\overline D^2
         \widetilde\nabla_\alpha
         +{1\over2}{\cal W}_B^\alpha\widetilde\nabla_\alpha
   \right)^{ab}V_Q^b.
\eqn\threexfour
$$

\section{BRST transformation}
To make the BRST symmetry in the unregularized theory manifest, we
adapt the following ghost-gauge fixing action:
$$
\eqalign{
   S'&=i\delta_{\rm BRST}
   {1\over T(R)}\int d^8z\,
   \tr(e^{-V_B}c^{\prime\dagger}e^{V_B}+c^\prime)V_Q
\cr
   &\qquad+{1\over T(R)}\int d^6z\,\tr Bf
   +{1\over T(R)}\int d^6\overline z\,\tr B^\dagger f^\dagger
\cr
   &\qquad-{2\xi\over T(R)}\int d^8z\,\tr e^{-V_B}f^\dagger e^{V_B}f
   -{2\xi\over T(R)}\int d^8z\,\tr e^{-V_B}b^\dagger e^{V_B}b,
\cr
}
\eqn\threexfive
$$
where we have introduced the Nakanishi-Lautrup (NL) superfield~$B$,
which is a Grassman-even chiral superfield,
$\overline D_{\dot\alpha}B=0$. The BRST transformation of the gauge
superfield is defined as usual by replacing the gauge
parameter~$\Lambda$ of the infinitesimal quantum field transformation
by the ghost superfield, $\Lambda\rightarrow c$ (see~(2.5) of~[\HAY]):
$$
\eqalign{
   &\delta_{\rm BRST}V_B=0,\qquad
   \delta_{\rm BRST}e^{V_Q}
   =i(e^{V_Q}c-e^{-V_B}c^\dagger e^{V_B}e^{V_Q}),
\cr
   &\delta_{\rm BRST}V_Q=
   i{\cal L}_{V_Q/2}\cdot\left[(c+e^{-V_B}c^\dagger e^{V_B})
   +\coth({\cal L}_{V_Q/2})\cdot(c-e^{-V_B}c^\dagger e^{V_B})\right],
\cr
   &\delta_{\rm BRST}\Phi_B=-ic\Phi_B,\qquad
    \delta_{\rm BRST}\Phi_Q=-ic\Phi_Q.
\cr
}
\eqn\threexsix
$$
Note that $\delta_{\rm BRST}$~is Grassman-odd in our convention. Then
the nilpotency of the BRST transformation determines the
transformation of the Faddeev-Popov (FP) ghost superfield:
$$
   \delta_{\rm BRST}c=-icc,\qquad
   \delta_{\rm BRST}c^\dagger=-ic^\dagger c^\dagger.
\eqn\threexseven
$$
The BRST transformation of the anti-ghost is defined to be the NL
superfield:
$$
\eqalign{
   &\delta_{\rm BRST}c^\prime=-iB,\qquad
    \delta_{\rm BRST}c^{\prime\dagger}=-iB^\dagger,
\cr
   &\delta_{\rm BRST}B=0,\qquad\delta_{\rm BRST}B^\dagger=0.
\cr
}
\eqn\threexeight
$$
We regard other fields, the gauge averaging function~$f$ and the NK
ghost~$b$, as BRST scalars, i.e.,
$\delta_{\rm BRST}f=\delta_{\rm BRST}b=0$.

The action~$S'$ \threexfive\ is equivalent to the conventional form of
the ghost-gauge fixing action in the superfield background field
method, \onexthree. This can easily be verified by eliminating the NL
field~$B$ from~\threexfive, using the equation of motion
$$
   \overline D^2V_Q=4f,
   \qquad{\cal D}^2V_Q=4e^{-V_B}f^\dagger e^{V_B}.
\eqn\threexnine
$$
Another equivalent form of~$S'$ is obtained by first integrating out
$f$ and~$f^\dagger$ and simultaneously $b$ and~$b^\dagger$
in~\threexfive. This integration may be performed by shifting the
variables as\foot{%
As was noted in Ref.~[\HAY], the combination in the denominator of
these expressions has to be interpreted as a representation of
$$
\eqalign{
   &\widetilde\nabla^2\overline D^2\leftrightarrow
   16\widetilde\nabla^m\widetilde\nabla_m
   -8{\cal W}_B^\alpha\widetilde\nabla_\alpha
   -4({\cal D}^\alpha{\cal W}_{B\alpha}),
\cr
   &\overline D^2\widetilde\nabla^2\leftrightarrow
   16\widetilde\nabla^m\widetilde\nabla_m
   +8\overline{\cal W}_{B\dot\alpha}^\prime\overline D^{\dot\alpha}
   +4(\overline D_{\dot\alpha}\overline{\cal W}_B^{\prime\alpha}),
\cr
}
$$
respectively.}
$$
   f^a\rightarrow
   f^a-{2\over\xi}\left(
   \overline D^2{1\over\widetilde\nabla^2\overline D^2}
   e^{-{\cal V}_B}\right)^{ab}B^{\dagger b},\qquad
   f^{\dagger a}\rightarrow
   f^{\dagger a}-{2\over\xi}\left(e^{{\cal V}_B}
   \widetilde\nabla^2{1\over\overline D^2\widetilde\nabla^2}
   \right)^{ab}B^b.
\eqn\threexten
$$
Then the Gaussian integral over~$f$ is precisely cancelled by the
integral of the NK ghost~$b$. After performing these integrations, the
ghost-gauge fixing term~$S'$~\threexfive\ is effectively replaced
by~$S'\rightarrow S_{\rm FP}+S_{\rm NL}$, where
$$
   S_{\rm FP}\equiv
   i\delta_{\rm BRST}\int d^8z\,V_Q^a
   \left[(e^{-{\cal V}_B})^{ab}c^{\prime\dagger b}
         +c^{\prime a}\right],
\eqn\threexeleven
$$
and
$$
   S_{\rm NL}\equiv{8\over\xi}\int d^8z\,B^a
   \left({1\over\widetilde\nabla^2\overline D^2}e^{-{\cal V}_B}
   \right)^{ab}B^{\dagger b}.
\eqn\threextwelve
$$
If one further integrates over the NL field~$B$, the conventional
gauge fixing term, $-(\xi/8)\int d^8z\,(\overline D^2V_Q)^a
(\widetilde\nabla^2 V_Q)^a$, again results. Interestingly, the
{\it kinetic operator\/} of~$B$,
$\overline D^2/(\widetilde\nabla^2\overline D^2)e^{-{\cal V}_B}D^2$,
is the {\it propagator\/} of the NK ghost. The Gaussian integration
of the bosonic NL field thus simulates the effect of the fermionic NK
ghost; this should be the case because \threexeleven\
plus~\threextwelve\ must be equivalent to~\onexthree. In what follows,
we use~$S_{\rm FP}+S_{\rm NL}$ as the ghost-gauge fixing action
because it exhibits the manifest BRST symmetry.

\section{Slavnov-Taylor identity}
As is well-known, the ST identities can be derived as expectation
values of a BRST exact expression. Our first example, which is
relevant to the discussion in the next section, is
$$
   \VEV{\VEV{T^*\delta_{\rm BRST}[c^{\prime a}(z)B^{\dagger b}(z')]}}
   =0,
\eqn\threexthirteen
$$
or
$$
   \VEV{\VEV{T^*B^a(z)B^{\dagger b}(z')}}=0.
\eqn\threexfourteen
$$
In these expressions, the double bracket~$\VEV{\VEV{\cdots}}$ implies
the would-be ``full'' expectation value, in which the BRST symmetry
is supposed to be exact. Similarly, we have
$$
   \VEV{\VEV{T^*\delta_{\rm BRST}[c^{\prime a}(z)V_Q^b(z')]}}=0,
\eqn\threexfifteen
$$
or
$$
\eqalign{
   &\VEV{\VEV{T^*B^a(z)V_Q^b(z')]}}
\cr
   &=-\VEV{\VEV{T^*c^{\prime a}(z)
   \left[
   {{\cal V}_Q\over2}\left(\coth{{\cal V}_Q\over2}+1\right)
   \right]^{bc}c^c(z')}}
\cr
   &\qquad
   +\VEV{\VEV{T^*c^{\prime a}(z)
   \left[
   {{\cal V}_Q\over2}\left(\coth{{\cal V}_Q\over2}-1\right)
   e^{-{\cal V}_B}\right]^{bc}c^{\dagger c}(z')}}.
\cr
}
\eqn\threexsixteen
$$
As mentioned above, the BRST symmetry is not manifest in our
regularization scheme. Therefore, the validity of the above relations
must be verified order by order. At present, we do not have a general
comment on this point. However, at least at the tree level and under
the classical equation of motion of the background field, there exists
a natural extension of our prescription which ensures the above
relations.

To examine the ST identities, we must first note that the
two-point functions in~\threexfourteen\ and~\threexsixteen\ are not
1PI, and thus the tadpole vertex $S_1^{\rm gauge}$~\threexthree\ also
contributes through, say, $S_{T3}$. Without taking into account these
tadpole contributions, the ST identities do not hold even in the
unregularized theory. This is related to the fact that each term in
the expansion $S^{\rm gauge}=S_0^{\rm gauge}+S_1^{\rm gauge}
+S_2^{\rm gauge}+\cdots$ is not individually invariant under the BRST
transformation~\threexsix. The tadpole contributions make a general
analysis of the ST identities in our scheme complicated.
 
Fortunately, for the application in the next section, a great
simplification occurs. We shall use the ST identities to conclude
that BRST exact composite operators vanish at the one-loop level.
These composite operators are defined by forming a one-loop diagram by
the (modified) two-point functions in~\threexfourteen\
and~\threexsixteen. Therefore even if the tadpole vertex~$S_1$ is
attached to the two-point functions, we can use the classical
equations of motion for $V_B$ and~$\Phi_B$, because we will work
with the one-loop approximation. Moreover, we will assume~$\Phi_B=0$
in the next section. This implies
${\cal D}^\alpha{\cal W}_{B\alpha}=0$ under the classical equation of
motion. Therefore the tadpole contributions arising
{}from~$S_1^{\rm gauge}$~\threexthree\ can be neglected because they
are proportional to~${\cal D}^\alpha{\cal W}_{B\alpha}$. When the
tadpole contribution can be neglected, the two-point functions in
\threexfourteen\ and~\threexsixteen\ at the tree level are given by
the propagators obtained by diagonalizing
$S_2^{\rm gauge}+S_{\rm FP}+S_{\rm NL}$. The Schwinger-Dyson
equations corresponding to~$S_2^{\rm gauge}+S_{\rm FP}+S_{\rm NL}$ are
given by
$$
\eqalign{
   &\VEV{T^*B^a(z)V_Q^b(z')}
   =-{\xi\over8}(\overline D^2\widetilde\nabla^2)^{ac}
   \VEV{T^*V_Q^c(z)V_Q^b(z')},
\cr
   &\VEV{T^*B^{\dagger a}(z)V_Q^b(z')}
   =-{\xi\over8}(e^{{\cal V}_B}\widetilde\nabla^2\overline D^2)^{ac}
   \VEV{T^*V_Q^c(z)V_Q^b(z')},
\cr
   &\VEV{T^*B^a(z)B^{\dagger b}(z')}
   =-{\xi\over8}
   (e^{{\cal V}_B'}\widetilde\nabla'^2\overline D'^2)^{bc}
   \Bigl[\VEV{T^*B^a(z)V_Q^c(z')}-i\delta^{ca}\delta(z-z')\Bigr],
\cr
}
\eqn\threexseventeen
$$
(the operator~$e^{{\cal V}_B'}\widetilde\nabla'^2\overline D'^2$ acts
on the $z'$-variable). In particular, these relations lead
to~$\VEV{T^*B^a(z)B^{\dagger b}(z')}=0$ at the unregularized level.

In implementing the regularization, we must specify how to modify
the original propagators. As a natural extension of the prescription
of~\S~1, we regard the first two relations of~\threexseventeen\
as the defining relations of $\VEV{T^*B^a(z)V_Q^b(z')}$
and~$\VEV{T^*B^{\dagger a}(z)V_Q^b(z')}$. Namely, on the right-hand
side of~\threexseventeen, the propagator of gauge superfield is
given by the modified one~\onexeight. On the other hand, we
{\it define\/} $\VEV{T^*B^a(z)B^{\dagger b}(z')}=0$ even in the
regularized theory.

It is obvious that the
prescription~$\VEV{T^*B^a(z)B^{\dagger b}(z')}=0$ is consistent
with~\threexfourteen\ at the tree level when~$S_1^{\rm gauge}=0$. On
the other hand, \threexsixteen\ at the tree level reads (by
suppressing quantum fields~$V_Q$),
$$
\eqalign{
   &\VEV{T^*B^a(z)V_Q^b(z')}
   =\left[e^{-{\cal V}_B}(z')\right]^{bc}
   \VEV{T^*c^{\prime a}(z)c^{\dagger c}(z')}
\cr
   &={i\over16}\Biggl[
   f\left([-\widetilde\nabla^m\widetilde\nabla_m
      +W_B^\alpha\widetilde\nabla_\alpha/2
      +({\cal D}^\alpha{\cal W}_{B\alpha})/4]/\Lambda^2\right)
\cr
   &\qquad\qquad\qquad\qquad
   \times\overline D^2{1\over\widetilde\nabla^m\widetilde\nabla_m
      -W_B^\alpha\widetilde\nabla_\alpha/2
      -({\cal D}^\alpha{\cal W}_{B\alpha})/4}
   \widetilde\nabla^2\Biggr]^{ab}\delta(z-z'),
\cr
}
\eqn\threexeighteen
$$
where the last expression is the modified ghost propagator~\onexnine\
in an un-abbreviated form~[\HAY]. At first glance, the left-hand side
of~\threexeighteen, i.e., \threexseventeen\ with~\onexeight, does not
seem to be identical to the right-hand side; in particular, the gauge
parameter~$\xi$ must disappear in the expression. To see that they
are in fact identical, when $S_1^{\rm gauge}=0$ or equivalently
when~${\cal D}^\alpha{\cal W}_{B\alpha}=0$, we first recall the
original form of the quadratic action of the gauge sector, the first
line of~\onexfour. Then, the first line of~\threexseventeen\ gives
$$
\eqalign{
   &\VEV{T^*B^a(z)V_Q^b(z')}
\cr
   &={i\over16}\left[\overline D^2\widetilde\nabla^2
   f(-\overline D^2\widetilde\nabla^2/16\Lambda^2+R/\xi\Lambda^2)
   {1\over\overline D^2\widetilde\nabla^2/16-R/\xi}
   \right]^{ab}\delta(z-z'),
\cr
}
\eqn\threexnineteen
$$
where the combination~$R$ has been defined by
$$
\eqalign{
   R&={1\over8}\widetilde\nabla^\alpha\overline D^2
         \widetilde\nabla_\alpha
     +{1\over2}{\cal W}_B^\alpha\widetilde\nabla_\alpha
     +{1\over4}{\cal D}^\alpha{\cal W}_{B\alpha}
     -{\xi\over16}\widetilde\nabla^2\overline D^2
\cr
   &={1\over8}\widetilde\nabla^\alpha\overline D^2
         \widetilde\nabla_\alpha
     +{1\over2}\widetilde\nabla^\alpha{\cal W}_{B\alpha}
     -{\xi\over16}\widetilde\nabla^2\overline D^2,\qquad
   \hbox{when ${\cal D}^\alpha{\cal W}_{B\alpha}\equiv
   \{\widetilde\nabla^\alpha,{\cal W}_{B\alpha}\}=0$}.
\cr
}
\eqn\threextwenty
$$
Since $\widetilde\nabla^2R=0$, one can eliminate $R$ in the
regularization factor~$f(t)$ of~\threexnineteen, and then
$\overline D^2\widetilde\nabla^2$ can be exchanged with the
regularization factor. Then, by noting the identity
$$
   \overline D^2\widetilde\nabla^2
   +\widetilde\nabla^2\overline D^2
   -2\overline D_{\dot\alpha}\widetilde\nabla^2
   \overline D^{\dot\alpha}
   =16\widetilde\nabla^m\widetilde\nabla_m
   -8{\cal W}_B^\alpha\widetilde\nabla_\alpha
   -4({\cal D}^\alpha {\cal W}_{B\alpha}),
\eqn\threextwentyone
$$
we have
$$
\eqalign{
   &\VEV{T^*B^a(z)V_Q^b(z')}
\cr
   &={i\over16}\Biggl[
   f\left([-\widetilde\nabla^m\widetilde\nabla_m
      +W_B^\alpha\widetilde\nabla_\alpha/2
      +({\cal D}^\alpha{\cal W}_{B\alpha})/4]/\Lambda^2\right)
\cr
   &\qquad\qquad\qquad\qquad\qquad\qquad\qquad\quad
   \times\overline D^2\widetilde\nabla^2
   {1\over\overline D^2\widetilde\nabla^2/16-R/\xi}
   \Biggr]^{ab}\delta(z-z').
\cr
}
\eqn\threextwentytwo
$$
Hence, by noting the identity~(3.8) of Ref.~[\HAY],
$$
   \overline D^2=
   \overline D^2
   {1\over\widetilde\nabla^m\widetilde\nabla_m
    -{\cal W}_B^\alpha\widetilde\nabla_\alpha/2
    -({\cal D}^\alpha{\cal W}_{B\alpha})/4}
    {\widetilde\nabla^2\overline D^2\over16},
\eqn\threextwentythree
$$
we realize that the last factor in~\threextwentytwo\ can be written as
$$
\eqalign{
   &\overline D^2\widetilde\nabla^2
   {1\over\overline D^2\widetilde\nabla^2/16-R/\xi}
\cr
   &=\overline D^2
   {1\over\widetilde\nabla^m\widetilde\nabla_m
    -{\cal W}_B^\alpha\widetilde\nabla_\alpha/2
    -({\cal D}^\alpha{\cal W}_{B\alpha})/4}
    {\widetilde\nabla^2\overline D^2\over16}
   \widetilde\nabla^2
   {1\over\overline D^2\widetilde\nabla^2/16-R/\xi}
\cr
   &=\overline D^2
   {1\over\widetilde\nabla^m\widetilde\nabla_m
    -{\cal W}_B^\alpha\widetilde\nabla_\alpha/2
    -({\cal D}^\alpha{\cal W}_{B\alpha})/4}
   \widetilde\nabla^2
   {\overline D^2\widetilde\nabla^2/16-R/\xi
    \over\overline D^2\widetilde\nabla^2/16-R/\xi}.
\cr
}
\eqn\threextwentyfour
$$
By substituting this into~\threextwentytwo, we see that
\threexnineteen~is in fact identical to~\threexeighteen. Therefore,
with the above prescription, the ST identity~\threexsixteen\
holds to at least the tree level under the classical equations of
motion, ${\cal D}^\alpha{\cal W}_{B\alpha}=0$. With these
understandings, we use the ST identities~\threexfourteen\ at the
tree level and~\threexeighteen\ in the next section.

\chapter{Superconformal anomaly in the gauge sector}
In the remainder of this paper, we describe a one-loop evaluation of
the superconformal anomaly in the gauge multiplet sector. This
problem was also postponed in Ref.~[\HAY]. Although the one-loop
result is well known~[\LUK--\MEH], our formulation allows a
{\it direct} calculation, relying neither on a supersymmetry--gauge
symmetry argument nor on the connection to the $\beta$-function. (Our
calculation is in spirit quite similar to that of Ref.~[\MAR].) We
also clarify the underlying BRST structure; the conventional form of
the superconformal anomaly accompanies BRST exact pieces that, due to
the Slavnov-Taylor identities in the previous section, eventually
vanish.

The superconformal anomaly is the spinor divergence of the
superconformal current,
$\overline D^{\dot\alpha}\VEV{R_{\alpha\dot\alpha}(z)}$. The
classical form of the superconformal current is given by~[\FER]
$$
   R_{\alpha\dot\alpha}
   =-{2\over T(R)}\tr W_\alpha e^{-V}\overline W_{\dot\alpha}e^V.
\eqn\fourxone
$$
Basically, what we have to do here is to construct an expansion of
this expression in powers of~$V_Q$ to~$O(V_Q^2)$ (the one-loop
approximation) and a classification of various terms in
$\overline D^{\dot\alpha}\VEV{R_{\alpha\dot\alpha}(z)}$ according to
their nature: the explicit breaking of superconformal symmetry
due to the gauge fixing, vanishing terms under the equation of motion
(see below), and the intrinsic quantum anomaly. This approach was
adopted in Ref.~[\HAY] for a computation of the superconformal anomaly
arising from the chiral matter's loop. However, the structure of the
direct expansion of~\fourxone\ is complicated, and it seems difficult
to directly perform such a classification. Therefore, we will adopt a
different strategy.

We start with the observation made by Shizuya~[\SHIZ] that the
superconformal current~\fourxone\ may be regarded as the Noether
current corresponding to a superfield variation,
$$
   \Delta e^V=-\Omega^\alpha e^VW_\alpha
   -\overline\Omega_{\dot\alpha}\overline W^{\dot\alpha}e^V,
\eqn\fourxtwo
$$
where $\Omega^\alpha$~is an infinitesimal Grassman-odd parameter. In
fact, one can easily verify that
$$
   \Delta S^{\rm gauge}=\int d^8z\,
   \left(-{1\over2}\Omega^\alpha
   \overline D^{\dot\alpha}R_{\alpha\dot\alpha}
   +{1\over2}\overline\Omega^{\dot\alpha}
   D^\alpha R_{\alpha\dot\alpha}\right),
\eqn\fourxthree
$$
by using the reality constraint on~$W_\alpha$.

To utilize the relation~\fourxthree\ in the background field method,
we split~$\Delta e^V$~\fourxtwo\ into variations of~$V_B$ and
of~$V_Q$. This splitting is of course not unique. Therefore, we can
impose the condition that a variation of~$V_B$ depends only on~$V_B$ and
a variation of~$V_Q$ is $O(V_Q)$. By writing~$\Delta
=\Delta_B+\Delta_Q$, this condition uniquely specifies\foot{%
In the following, we explicitly write only the $\Omega^\alpha$-parts.}
$$
\eqalign{
   &\Delta_Be^{V_B}\equiv-\Omega^\alpha e^{V_B}W_{B\alpha},\qquad
   \Delta_B V_Q\equiv0,
\cr
   &\Delta_QV_B\equiv0,\qquad
   \Delta_Q e^{V_Q}\equiv
   \Omega^\alpha
   \left[-e^{V_Q}(W_\alpha-W_{B\alpha})+[W_{B\alpha},e^{V_Q}]\right],
\cr
   &{\Delta_B\atop\Delta_Q}\biggr\}(\hbox{other fields})
   \equiv0,
\cr
}
\eqn\fourxfour
$$
and thus
$$
   \Delta_Q V_Q^a=\Omega^\alpha
   \left({1\over4}\overline D^2\widetilde\nabla_\alpha
         +{\cal W}_{B\alpha}\right)^{ab}V_Q^b+O(V_Q^2).
\eqn\fourxfive
$$

Now, as is clear from~\fourxthree, the calculation of the
superconformal anomaly is equivalent to the evaluation of the 1PI part
of
$$
   \VEV{\Delta S^{\rm gauge}}
   =\VEV{\Delta_BS^{\rm gauge}}+\VEV{\Delta_QS^{\rm gauge}}.
\eqn\fourxsix
$$
At this stage, we should note the following fact. In classical
theory, a conservation law follows from the classical equation of
motion. In quantum theory, the classical equation of motion is
replaced by the ``quantum'' equation of motion,
$\delta{\mit\Gamma}[V_B]/\delta V_B^a=0$, where~${\mit\Gamma}[V_B]$
is the effective action in the background field method.\foot{%
The equivalence of the effective action in the background field
method and that of the conventional formalism is proven in
Ref.~[\ABBO] for non-supersymmetric gauge theories. It is argued
in Ref.~[\GAT] for the supersymmetric case.}
Without using this equation to an accuracy consistent with the
treatment, the conservation law does not follow even if there is no
quantum anomaly nor explicit breaking. An example of such a
``quantum'' equation of motion is found in~(5.18) of Ref.~[\HAY].
Therefore, we should consider the superconformal
anomaly~$\overline D^{\dot\alpha}\VEV{R_{\alpha\dot\alpha}(z)}$ under
the quantum equation of motion. In our present problem,
$\delta{\mit\Gamma}[V_B]/\delta V_B^a=0$~implies, in particular, at
the one-loop level,
$$
   \VEV{\Delta_BS^{\rm gauge}}
   =-\VEV{\Delta_BS_{\rm FP}}-\VEV{\Delta_BS_{\rm NL}},
\eqn\fourxseven
$$
because of our prescription for the one-loop effective
action~\onexthirteen. Recall that
$S^{\rm gauge}+S_{\rm FP}+S_{\rm NL}$ is the total action (see
\threexone, \threexeleven\ and~\threextwelve).

On the other hand, we rewrite the quantum variation part of~\fourxsix\
as
$$
   \VEV{\Delta_QS^{\rm gauge}}
   =\VEV{\Delta_Q(S^{\rm gauge}+S_{\rm FP})}
   -\VEV{\Delta_QS_{\rm FP}}.
\eqn\fourxeight
$$
Therefore we have, from \fourxsix--\fourxeight,
$$
   \VEV{\Delta S^{\rm gauge}}
   =\VEV{\Delta_Q(S^{\rm gauge}+S_{\rm FP})}
   -\VEV{(\Delta_B+\Delta_Q)S_{\rm FP}}-\VEV{\Delta_BS_{\rm NL}},
\eqn\fourxnine
$$
under the quantum equation of motion of~$V_B$.

In what follows, we show that the first
piece~$\VEV{\Delta_Q(S^{\rm gauge}+S_{\rm FP})}$ in~\fourxnine\ gives
rise to the quantum anomaly and that the remaining pieces
$(\Delta_B+\Delta_Q)S_{\rm FP}$ and~$\Delta_BS_{\rm NL}$ are BRST
{\it exact}. This fact is obvious for the last
term~$\Delta_BS_{\rm NL}$ because the NL field~$B$ is the BRST
transformation of the anti-ghost, \threexeight\ (note $\Delta_BB=0$).
The BRST exactness
of~$(\Delta_B+\Delta_Q)S_{\rm FP}$ follows from the remarkable
property of $\Delta_B$ and~$\Delta_Q$ that their {\it sum\/} commutes
with the BRST transformation~$\delta_{\rm BRST}$. In fact, one can
confirm
$$
\eqalign{
   &(\Delta_B+\Delta_Q)\delta_{\rm BRST}e^{V_Q}
\cr
   &=i\Omega^\alpha\left[
   (-e^{V_Q}W_\alpha+W_{B\alpha}e^{V_Q})c
   -e^{-V_B}c^\dagger e^{V_B}e^{V_Q}W_\alpha
   -W_{B\alpha}e^{-V_B}c^\dagger e^{V_B}e^{V_Q}\right]
\cr
   &=\delta_{\rm BRST}(\Delta_B+\Delta_Q)e^{V_Q},
\cr
}
\eqn\fourxten
$$
and
$$
   (\Delta_B+\Delta_Q)\delta_{\rm BRST}(\hbox{other fields})=0
   =\delta_{\rm BRST}(\Delta_B+\Delta_Q)(\hbox{other fields}).
\eqn\fourxeleven
$$
Therefore these two operations commute
$[(\Delta_B+\Delta_Q),\delta_{\rm BRST}]=0$ on all the fields. As a
result of this property, we see
$$
\eqalign{
   &\VEV{(\Delta_B+\Delta_Q)S_{\rm FP}}
\cr
   &=\VEV{i\delta_{\rm BRST}
   \int d^8z\,(\Delta_B+\Delta_Q)V_Q^a
   \left[(e^{-{\cal V}_B})^{ab}c^{\prime\dagger b}
   +c^{\prime a}\right]}
\cr
   &=\int d^8z\,\Omega^\alpha
\cr
   &\times
   \VEV{i\delta_{\rm BRST}
   \left[
   c^{\prime\dagger a}
   \left({1\over4}e^{{\cal V}_B}
   \overline D^2\widetilde\nabla_\alpha\right)^{ab}V_Q^b
   +c^{\prime a}
   \left({1\over4}\overline D^2\widetilde\nabla_\alpha
   +{\cal W}_{B\alpha}\right)^{ab}V_Q^b+O(V_Q^2)\right]}.
\cr
}
\eqn\fourxtwelve
$$
If the theory (including the regularization) is BRST invariant, the
expectation value of a BRST exact piece must vanish in a BRST
invariant state. Although this is not manifest in our regularization
scheme, we have shown in the previous section that the tree level ST
identities, $\VEV{T^*B^a(z)B^{\dagger b}(z')}=0$ and~\threexeighteen,
hold with an appropriate prescription under the classical equation
of motion, ${\cal D}^\alpha{\cal W}_{B\alpha}=0$ (we are assuming
$\Phi_B=0$). As a result, by defining the composite operators in
$\VEV{\Delta_BS_{\rm NL}}$ and~\fourxtwelve\ by connecting the
quantum fields by modified propagators, according to~\onexfourteen,
we can safely neglect the BRST exact pieces, at least at the one-loop
level.

Having observed that the last two terms in~\fourxnine\ are BRST exact
and can be neglected even in our present formulation, let us return
to the first term on the right-hand side of~\fourxnine. To one-loop
accuracy, there are three terms which may contribute to the 1PI part
of~$\VEV{\Delta_Q(S^{\rm gauge}+S_{\rm FP})}$:
(i)~an $O(V_Q^2)$~term in~$\Delta_QS_1^{\rm gauge}$,
(ii)~an $O(V_Q^2)$~term in~$\Delta_QS_2^{\rm gauge}$, and
(iii)~an $O(V_QB)$~term in~$\Delta_QS_{\rm FP}$. However, the first
contribution~(i) is proportional to
${\cal D}^\alpha{\cal W}_{B\alpha}$ (see~\threexthree), which is
already a one-loop order quantity under the quantum equation of
motion (recall that the classical equation of motion
is~${\cal D}^\alpha{\cal W}_{B\alpha}=0$). As a result, (i)~is a
higher-order quantity and can be neglected. To evaluate (ii)
and~(iii), the expression~\fourxfive\ is sufficient. Therefore, from
\threexfour\ and~\threexeleven, we have
$$
\eqalign{
   &\VEV{\Delta_Q(S^{\rm gauge}+S_{\rm FP})}
\cr
   &=\int d^8z\,2\Omega^\alpha
   \VEV{
   \left({1\over4}\overline D^2\widetilde\nabla_\alpha
         +{\cal W}_{B\alpha}\right)^{ab}V_Q^b(z)
   \left({1\over8}\widetilde\nabla^\beta\overline D^2
         \widetilde\nabla_\beta
         +{1\over2}{\cal W}_B^\beta\widetilde\nabla_\beta
   \right)^{ac}V_Q^c(z)}
\cr
   &\qquad+\int d^8z\,\Omega^\alpha
   \VEV{
   \left({1\over4}\overline D^2\widetilde\nabla_\alpha
         +{\cal W}_{B\alpha}\right)^{ab}V_Q^b(z)
   \left[(e^{-{\cal V}_B})^{ac}B^{\dagger c}(z)
         +B^a(z)\right]}
\cr
   &=-{i\over4}\int d^8z\,\Omega^\alpha
   \lim_{z'\rightarrow z}
   \biggl[
   f\left((-\widetilde\nabla^m\widetilde\nabla_m
   +{\cal W}_B^\beta\widetilde\nabla_\beta/2
   -\overline{\cal W}_{B\dot\beta}^\prime
    \overline D^{\dot\beta}/2)/\Lambda^2\right)
\cr
   &\qquad\qquad\qquad\qquad\qquad\qquad\qquad\qquad\qquad\qquad
   \quad
   \times(\overline D^2\widetilde\nabla_\alpha-4{\cal W}_{B\alpha})
   \biggr]^{aa}\delta(z-z'),
\cr
}
\eqn\fourxthirteen
$$
where, in the last expression, we have used the first two lines
of~\threexseventeen\ and~\onexeight\ in the super-Fermi-Feynman
gauge~$\xi=1$. Since \fourxthirteen~vanishes in the absence of the
regularization factor~$f(t)$, we realize that this is in fact the
quantum anomaly (i.e., a naive interchange of the equal-point limit
and $\Lambda\rightarrow\infty$~limit gives an incorrect result). The
computation of the last expression proceeds in an almost identical
way as that in the Appendix of Ref.~[\HAY] and we have\foot{%
The expansion of the regulating factor~$f(t)$ becomes considerably
simpler if one notes that the whole expression is a superfield by
construction, and thus explicit dependences on~$\theta^\alpha$ and
$\overline\theta_\alpha$ must vanish eventually.}
(the $-4{\cal W}_{B\alpha}$~part does not contribute)
$$
\eqalign{
   &\lim_{z'\rightarrow z}
   \left[
   f\left((-\widetilde\nabla^m\widetilde\nabla_m
   +{\cal W}_B^\beta\widetilde\nabla_\beta/2
   -\overline{\cal W}_{B\dot\beta}^\prime
   \overline D^{\dot\beta}/2)/\Lambda^2\right)
   \overline D^2\widetilde\nabla_\alpha\right]^{aa}
   \delta(z-z')
\cr
   &\Lambdato-{i\over16\pi^2}
   ({\cal W}_B^\beta{\cal D}_\beta{\cal W}_{B\alpha})^{aa}
\cr
   &={i\over32\pi^2}
   \left[D_\alpha({\cal W}_B^\beta{\cal W}_{B\beta})^{aa}
   +2({\cal W}_{B\alpha}{\cal D}^\beta{\cal W}_{B\beta})^{aa}\right]
\cr
   &={i\over32\pi^2}{C_2(G)\over T(R)}
   D_\alpha\tr W_B^\beta W_{B\beta},\qquad
   \hbox{when ${\cal D}^\beta{\cal W}_{B\beta}=0$}.
\cr
}
\eqn\fourxfourteen
$$

Finally, by combining equations \fourxfourteen, \fourxthirteen,
\fourxnine\ and~\fourxthree, we obtain, to one-loop accuracy,
$$
   \overline D^{\dot\alpha}\VEV{R_{\alpha\dot\alpha}(z)}
   \Lambdato-{1\over64\pi^2}
   {C_2(G)\over T(R)}D_\alpha\tr W_B^\beta W_{B\beta}.
\eqn\fourxfifteen
$$
We emphasize that we did not need the expansion in~$V_B$ in deriving
this expression. Also, it is straightforward to include the effect of
the chiral multiplet in the calculation, at least when the background
chiral superfield is absent (i.e.\ when $\Phi_B=0$). By utilizing the
result of Ref.~[\HAY],\foot{%
Or, one may use the fact~[\SHIZ] that the superconformal current of
the chiral multiplet can be regarded as the Noether current
associated with the variation $\Delta\Phi=-\overline D^2
[\Omega^\alpha e^{-V}D_\alpha e^V\Phi
+(D^\alpha\Omega_\alpha)\Phi/3]/4$ and repeat the above procedure.}
we have
$$
\eqalign{
   &\overline D^{\dot\alpha}\VEV{R_{\alpha\dot\alpha}(z)}
   -2\VEV{\Phi_Q^Tm\nabla_\alpha\Phi_Q(z)}
   +{2\over3}D_\alpha\VEV{\Phi_Q^Tm\Phi_Q(z)}
\cr
   &\Lambdato
   -{1\over8\pi^2}
   \Biggl[\Lambda^2\int_0^\infty dt\,f(t)+{1\over6}\dA\Biggr]
   \tr W_{B\alpha}(z)
   -{3C_2(G)-T(R)\over 192\pi^2}{1\over T(R)}
   D_\alpha\tr W_B^\beta W_{B\beta}.
\cr
}
\eqn\fourxsixteen
$$

One of the advantages of our scheme is that it provides a
supersymmetric (background) gauge-invariant definition of the
superconformal current operator~[\HAY]. Therefore, from the
expression of the superconformal anomaly~\fourxsixteen, we can
immediately derive (quantum anomalous part of) the ``central
extension'' of $N=1$~supersymmetry algebra~[\DVA]; the procedure
in Ref.~[\HAY] gives (for $m=0$)
$$
\eqalign{
   \VEV{\{\overline Q_{\dot\alpha},\overline Q_{\dot\beta}\}}
   &={i\over2}\int d^3x\,
   \overline\sigma^{0\dot\gamma\gamma}
   (\varepsilon_{\dot\alpha\dot\beta}\overline D_{\dot\gamma}
   +2\varepsilon_{\dot\beta\dot\gamma}\overline D_{\dot\alpha})
   \left.\overline D^{\dot\delta}
   \VEV{R_{\gamma\dot\delta}(z)}\right|_{\theta=\overline\theta=0}
\cr
   &\Lambdato{1\over48\pi^2}[3C_2(G)-T(R)]
   \overline\sigma^{0i}_{\dot\alpha\dot\beta}\int d^3x\,\partial_i
   (\lambda_B^{\alpha a}\lambda_{B\alpha}^a).
\cr
}
\eqn\fourxseventeen
$$
If we retained the BRST exact pieces in~\fourxnine\ in the
superconformal anomaly, the central extension~\fourxseventeen\ also
acquires BRST exact pieces. It is interesting that this structure is
common to the result of Ref.~[\FUJIK], although we are using a
supersymmetric invariant gauge fixing term.

\chapter{Conclusion}
In this paper, we have reported on the analyses of several issues
postponed in Ref.~[\HAY], including the proof of the covariant gauge
current in anomaly free cases. The basic properties of our scheme
(except the BRST symmetry at higher orders) have now been clarified,
and it has been established that our scheme actually provides a gauge
invariant regularization of the effective action in the background
field method. On the practical side, our scheme passed many one-loop
tests performed in Ref.~[\HAY] and the present paper by reproducing
the correct results in a transparent manner. Therefore, we are now
ready to use our scheme in treating more realistic problems for which
a manifestly supersymmetric gauge covariant treatment is crucial.

We are grateful to Professor K. Fujikawa and Dr.~J. Hisano for
helpful comments. The work of H.S. is supported in part by the
Ministry of Education Grant-in-Aid Scientific Research, Nos.~09740187
and~10120201.

\refout
\bye